\documentclass{article}
\usepackage{spconf,amsmath,graphicx}
\usepackage{multirow,amssymb,hyperref}
\usepackage[numbers,sort&compress]{natbib}
\usepackage{booktabs,url}
\usepackage{psfrag,pstricks,float,listings,color}
\usepackage{algorithm,algpseudocode,booktabs}
\usepackage{multirow}
\usepackage{bm}

\usepackage[utf8]{inputenc}

\usepackage{mathtools}
\usepackage{commath}

\let\norm\undefined 
\DeclarePairedDelimiter\norm{\lVert}{\rVert}
\DeclarePairedDelimiter\floor{\lfloor}{\rfloor}

\title{IMPROVED LARGE-MARGIN SOFTMAX LOSS FOR SPEAKER DIARISATION}
\name{Y. Fathullah, C. Zhang and P. C. Woodland}
\address{Cambridge University Engineering Department, UK \\ \small{\texttt{\{yf286,cz277\}@cam.ac.uk, pcw@eng.cam.ac.uk}}\\}
\begin{document}
	\ninept
	\maketitle
	\begin{abstract}
		Speaker diarisation systems nowadays use embeddings generated from speech segments in a bottleneck layer, which are needed to be discriminative for unseen speakers. 
		It is well-known that large-margin training can improve the generalisation ability to unseen data, and its use in such open-set problems has been widespread. 
		Therefore, this paper introduces a general approach to the large-margin softmax loss without any approximations to improve the quality of speaker embeddings for diarisation. Furthermore, a novel and simple way to stabilise training, when large-margin softmax is used, is proposed. Finally, to combat the effect of overlapping speech, different training margins are used to reduce the negative effect overlapping speech has on creating discriminative embeddings. Experiments on the AMI meeting corpus show that the use of large-margin softmax significantly improves the speaker error rate (SER). By using all hyper parameters of the loss in a unified way, further improvements were achieved which reached a relative SER reduction of 24.6\% over the baseline. However, by training overlapping and single speaker speech samples with different margins, the best result was achieved, giving overall a 29.5\% SER reduction relative to the baseline.  
	\end{abstract}
	\section{Introduction} \label{sec:Introduction}
	
	\begin{figure*}[t]
		\centering
		\includegraphics[width=0.98\textwidth]{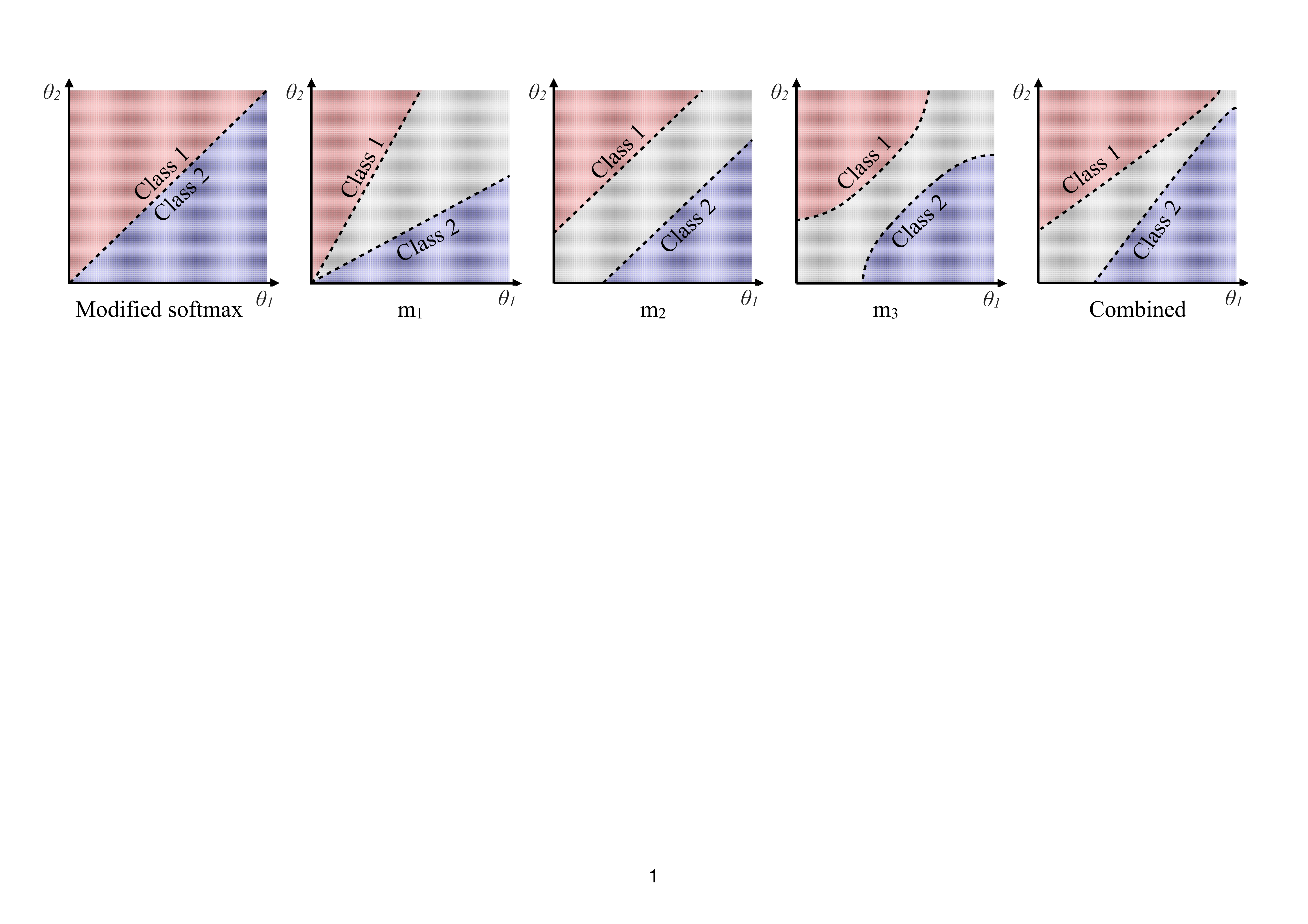}
		\vspace{-0.3cm}
		\caption{A visual comparison of decision boundaries/criteria between modified softmax loss and large-margin softmax. Red and blue represent the two classes and grey is the decision region. The decision region is comprised of the area between the two decision boundaries $\psi(\theta_1) = \cos(\theta_2)$ and $\psi(\theta_2) = \cos(\theta_1)$ (the dotted lines), which coincide for modified softmax. The combined version using all parameters $m_1$, $m_2$, $m_3$ can achieve new improved types of decision regions (using Eqn. (\ref{eq:gen_psi})) which are not possible by the previous setup.}
		\label{fig:decisionBoundaries}
	\end{figure*}
	
	The goal of speaker diarisation is to answer the question of ``who spoke when'' given a multi-speaker audio input. Generally, a speaker diarisation system partitions an audio stream into single speaker segments and identifies segments belonging to the same speakers. This is achieved through multiple stages using several models. The first stage identifies segments consisting of a single speaker, with the use of voice activity and speaker change point detection. The second stage extracts features from these segments, to represent the speakers, from now on referred to as \textit{speaker embeddings}. In the third and final stage, the embeddings are clustered. \cite{X. A. Miro et al, S. Tranter et al} 
	
	Over the past years with the introduction of deep learning, diarisation systems have greatly improved in performance \cite{E. Variani et al, G. Heigold et al, D. Garcia-Romero et al}. Going from the use of i-vectors to d-vectors, for representing a segment of speech for a single speaker, has been a contributing factor to progress \cite{S.H. Yella et al v3, M. McLaren et al}. The second factor has been the use of more sophisticated clustering processes, which are more appropriate for diarisation \cite{Q. Wang et al, S. H. Shum et al}. However, more work needs to be done to improve these systems. They need to, for example, be robust to noisy backgrounds, moving speakers and overlapping speech \cite{P. Cyrta et al}. Additionally, since speaker diarisation systems are used for open-set problems, in environments with unseen speakers, the embeddings need to be discriminative enough for a clustering or supervised process to be able to differentiate between unseen speakers \cite{Y. Liu et al, A. Zhang et al}.
	
	To achieve discriminative features that are good enough to generalise to unseen speakers, the compactness of a class, and distance to other classes should be maximised during training. This was first tackled by the triplet loss in \cite{C. Li et al, C. Zhang et al}, which explicitly maximised inter-class separability and intra-class compactness. It took three embeddings, and minimised the distance between the anchor and the positive embedding (belonging to the same class) and at the same time maximised the distance between the anchor and the negative embedding (which belong to different speakers). However, the triplet loss is sensitive to the data mining strategy and requires considerable efforts \cite{C.-Y. Wu et al}. Instead work has been done on improving the cross-entropy objective by using different ways of encouraging samples to their class centres (center-loss), or maximising distances between the class centres (minimum-hyperspherical energy) \cite{Y. Wen et al, W. Liu et al MHE}. Finally a large-margin softmax loss (L-Softmax) has been introduced, which can be seen as a generalisation of the cross-entropy loss \cite{W. Liu et al, W. Liu SphereFace et al, H. Wang et al, J. Deng et al}. Just as cross-entropy is computed by applying a softmax activation function to the activations of the last layer, the large-margin softmax loss first increases the angle between the embedding and the target class, therefore, reducing the target activation. Thereafter, it applies the softmax function. This can be viewed as the target class being separated from remaining classes allowing for a margin to be created. It can also be seen as the loss enforcing a stricter decision criteria compared to the cross-entropy loss, therefore, requiring the embedding to be closer to the target class centre, resulting in more compact classes.
	
	In this paper, a general version of the large-margin softmax loss is proposed and combined with spectral clustering, since it can make use of the angular nature of the embeddings \cite{Q. Wang et al}. The proposed general large-margin softmax loss (GLM-Softmax) builds on L-Softmax by adding flexibility and is needed to allow for a more discriminative model. This is because the loss, compared to cross-entropy, encourages more inter-class separability and intra-class compactness, at the cost of introducing hyper parameters \cite{W. Liu et al, Y. Liu et al}. In addition, GLM-Softmax will allow three different types of hyper parameters to be used in a unified way, simultaneously, to make use of the properties of each one. This paper will also combat the negative effects of overlapping speech by making the GLM-Softmax enforce a less strict decision criteria for these samples, instead using the same decision criteria as a cross-entropy loss with normalised weights and zero bias values (referred to as \textit{modified softmax loss}). 
	This will, in practice, reduce the effect of overlapping speech on the inter-class separability and intra-class compactness thus, leading to embeddings of higher quality. It has also been noted that large-margin softmax can be unstable and, therefore, a simpler way of achieving stability compared to the method used in \cite{Y. Liu et al, W. Liu et al, W. Liu SphereFace et al} is introduced.
	
	The remainder of this paper is structured as follows: Section \ref{sec:preliminaries} covers the original large-margin softmax. Section \ref{sec:proposed} describes the proposed general large-margin softmax. The experimental setup, results and conclusion can be found in Secs. \ref{sec:experimental}, \ref{sec:results} and \ref{sec:conclusion} respectively.
	
	\section{Preliminaries} \label{sec:preliminaries}
	
	The large-margin softmax approach arose from the demand to create discriminative features when generalising to open-set problems. It can either be motivated from how the loss modifies the target activation of the feature layer or how the decision boundaries (or criteria) of large-margin softmax and cross-entropy differ. The following subsections present both the original definition and a treatment of the decision boundary in order to explain the properties of the loss.
	
	\subsection{Definition} \label{sesc:definition}
	
	The L-Softmax \cite{W. Liu et al} loss was originally defined as:
	\begin{equation*}
	\mathcal{L} \hspace*{-0.2mm} = \hspace*{-0.2mm} - \hspace*{-0.2mm} \ln \hspace*{-0.2mm} \Bigg( \frac{e^{\norm{\mathbf{x}} \norm{\mathbf{w}_{t}} \psi(\theta_{t})} }{e^{\norm{\mathbf{x}} \norm{\mathbf{w}_{t}} \psi(\theta_{t})} \hspace*{-0.2mm} + \hspace*{-0.2mm} \sum_{c \neq t} e^{\norm{\mathbf{x}} \norm{\mathbf{w}_c}\cos(\theta_{c})}} \hspace*{-0.2mm} \Bigg)
	\end{equation*}
	where $\mathbf{x}$ is the speaker embedding, $\mathbf{w}_c$ is the weight vector associated with class $c$ (with $t$ being the target class) and $\theta_{c}$ is the angle between them.
	Here $\psi(\theta_t)$ is a monotonically non-increasing function that satisfies $\cos(\theta_t) \geqslant \psi(\theta_t)$ for $\theta_t \in [0, \pi]$. It can be seen as a generalised cosine function and was defined as: 
	\begin{equation} \label{eq:psi}
	\psi(\theta_t) = (-1)^k \cos(m\theta_t) - 2k, \medspace \medspace \medspace \theta_t \in [\frac{k\pi}{m}, \frac{(k + 1)\pi}{m}]
	\end{equation}
	where the integer $k \in [0, m-1]$ and $m$ is an integer due to the choice of implementation \cite{W. Liu et al, W. Liu SphereFace et al}. The use of $k$ ensures that $\psi(\cdot)$ is non-increasing, however, related work \cite{J. Deng et al, Y. Liu et al} has discarded $k$ and introduced multiple different parameters resulting in an approximated $\psi(\theta_t)$:
	\begin{equation} \label{eq:approxpsi}
	\psi(\theta_t) \approx \cos(m_1\theta_t + m_2) - m_3
	\end{equation}
	which also allows real values of $m_1$ \cite{J. Deng et al, Y. Liu et al}. 
	This approximation violates the conditions that $\psi(\cdot)$ needs to satisfy and will therefore have an impact on the models. Larger parameters values of $m_1$ and $m_2$ will lead to poorer approximations and ultimately to a limitation of performance. Additionally, it can be noted that large-margin losses utilising only $m_3$ are unaffected by this approximation, giving these models a possible advantage from an implementation viewpoint.
	
	\subsection{Decision Boundary} \label{ssec:decision}
	
	To offer insight into the properties large-margin softmax, such as why classes are separated and why it could be a harder loss function to optimise, the decision boundary can be studied. The cross-entropy decision boundary in the two class case satisfies:
	\begin{align*}
	\mathbf{x}^T(\mathbf{w}_1 - \mathbf{w}_2) + b_1 - b_2 & =  \\
	\norm{\mathbf{x}}(\norm{\mathbf{w}_1}\cos(\theta_1) - \norm{\mathbf{w}_2}\cos(\theta_2)) + b_1 - b_2 & = 0
	\end{align*}
	where $\theta_c$ denotes the angle between feature $\mathbf{x}$ and the weight vector for class $c$ $\mathbf{w}_c$. For modified softmax loss ($b_c = 0$, $\norm{\mathbf{w}_c} = 1$) the boundary satisfies: $\cos(\theta_1) - \cos(\theta_2) = 0$. The decision boundary is symmetric and the same irrespective of which class $\mathbf{x}$ belongs to. However, for large-margin softmax losses, if $\mathbf{x}$ belongs to class 1, the boundary is instead $\psi(\theta_1) - \cos(\theta_2) = 0$, which represents a stricter decision for class 1. This can be shown from the definition of $\psi(\cdot)$ that if a sample (belonging to class 1) is classified correctly in the strict case, it is correctly classified using cross-entropy since
	\begin{equation} \label{eq:hardInequality}
	\cos(\theta_1) \geqslant \psi(\theta_1) > \cos(\theta_2)
	\end{equation}
	Class 2 has its own unique boundary $\psi(\theta_2) - \cos(\theta_1) = 0$ and due to $\psi(\cdot)$ satisfying $\cos(\theta_2) \geqslant \psi(\theta_2)$ there will be a decision region in between the two boundaries separating the two classes, see Fig. \ref{fig:decisionBoundaries}. The inequality (\ref{eq:hardInequality}) above is also an indication of the family of L-Softmax being harder loss functions to optimise, and therefore, could lead to convergence issues, especially for difficult problems like speaker clustering with the presence of overlapping speech.
	
	\section{Proposed Approach} \label{sec:proposed}
	
	In this section, a unified approach to large-margin softmax using all parameters $m_1, m_2, m_3$ without any approximation is presented. Furthermore, it has been noted in many works that training can be unstable \cite{W. Liu et al, W. Liu SphereFace et al, Y. Liu et al} and therefore, a new way of achieving stability is introduced. Lastly, since overlapping speech can be present in the data, a novel method to reduce the effect of this factor on the speaker embeddings is suggested.
	
	\subsection{Introducing the General Large-Margin Softmax} \label{ssec:implementation}
	
	A definition for $\psi(\cdot)$ is first provided:
	\begin{equation} \label{eq:gen_psi}
	\psi(\theta_t) \coloneqq (-1)^k \cos(m_1\theta_t + m_2) - m_3 - 2k
	\end{equation}
	Instead of only having a multiplicative margin $m_1$ as in equation (\ref{eq:psi}), the $\psi(\cdot)$ in equation (\ref{eq:gen_psi}) combines the strengths of both (\ref{eq:psi}) and (\ref{eq:approxpsi}). This gives rise to more flexible decision boundaries (see Fig. \ref{fig:decisionBoundaries}) without any approximation (as in Eqn. (\ref{eq:approxpsi})), and will therefore, not limit performance in case large values of $m_1$ and $m_2$ are needed.
	
	Two sufficient conditions on $\psi(\theta_t)$ need to be satisfied for a proper decision margin to exist between classes: being monotonically non-increasing; and $\cos(\theta_t) \geqslant \psi(\theta_t)$ for $\theta_t \in [0, \pi]$. This leads to requiring
	\begin{equation*}
	\theta_t \in [\frac{k\pi - m_2}{m_1}, \frac{(k+1)\pi - m_2}{m_1}], \medspace\medspace k = 0, 1, ..., \floor{m_1} + 1
	\end{equation*}
	to ensure the conditions are met. When the parameters are used individually, the allowed values are $m_1 \geqslant 1$, $m_2 \in [0, \pi]$ and $m_3 \geqslant 0$, respectively. However, when using the parameters in combination, values such as $m_1 < 1$ or $m_3 < 0$ can be allowed if the other parameters are large enough to ensure $\cos(\theta_t) \geqslant \psi(\theta_t)$. 
	
	From an optimisation viewpoint, extending $m_1$ to non-integer values will lead to a higher computational cost since the multiple angle formula can no longer be used. Instead of using the already computed values of $\mathbf{x}^T\mathbf{w}_t$ from the forward propagation, trigonometric functions need to be used to compute the gradients.
	
	The gradients for $f(\mathbf{x}, \mathbf{w}_t) = \norm{\mathbf{x}}\norm{\mathbf{w}_t}\psi(\theta_t)$ become:
	\begin{align*}
	&\frac{\partial f(\mathbf{x}, \mathbf{w}_t)}{\partial \mathbf{x}} = \frac{\mathbf{x}}{\norm{\mathbf{x}}}\norm{\mathbf{w}_t}\psi(\theta_t) + \norm{\mathbf{x}}\norm{\mathbf{w}_t}\frac{\partial \psi(\theta_t)}{\partial \mathbf{x}}\\
	&\frac{\partial \psi(\theta_t)}{\partial \mathbf{x}} = m_1(-1)^k \frac{\sin(m_1\theta_t \hspace{-0.3mm}+\hspace{-0.3mm} m_2)}{\abs{\sin(\theta_t)}} \bigg(\frac{\mathbf{w}_t}{\norm{\mathbf{w}_t}\norm{\mathbf{x}}} \hspace{-0.2mm}-\hspace{-0.2mm} \frac{(\mathbf{w}_t^{\text T} \mathbf{x}) \mathbf{x}}{\norm{\mathbf{w}_t}\norm{\mathbf{x}}^3}\bigg)
	\end{align*}
	with symmetrical expressions for gradients with respect to $\mathbf{w}_t$. This, therefore, requires computing the angle $\theta_t$ by $\arccos(\cdot)$, thereafter finding $k$ before being able to find the gradients for the loss function.
	
	\subsection{Improving Training Stability for GLM-Softmax} \label{ssec:improving}
	
	As mentioned in Section \ref{sec:preliminaries}, the GLM-Softmax could be harder to optimise because it represents a stricter criteria compared to cross-entropy. Therefore, to stabilise training, a weighted average between $\psi(\theta_t)$ and $\cos(\theta_t)$ has been proposed and used by multiple authors \cite{W. Liu et al, W. Liu SphereFace et al, Y. Liu et al}. Instead, in this paper, it is suggested that a more natural way to stabilise training is to gradually increase the parameters to the desired value, after the $n$-th weight update step, using:
	\begin{equation} \label{eq:updateSchedule}
	m_i^{(n)} = m_i^{(n-1)} + \eta\,(m_i - m_i^{(n-1)}), \medspace \medspace m_i^{(0)} = \Big\{ \begin{array}{l}
	1, \thickspace \medspace i = 1 \\
	0, \thickspace \medspace i = 2, 3
	\end{array}
	\end{equation}
	Here, $m_i^{(0)}$ and $m_i$ represent the starting and final values of the GLM-softmax hyper parameters, and $\eta \in (0,1)$ controls the update step size of $m_i^{(n)}$.  The method used elsewhere \cite{W. Liu et al, W. Liu SphereFace et al, Y. Liu et al} requires setting the decay, start and final value for the parameter that controls the weighting, while the update schedule proposed in Eqn. (\ref{eq:updateSchedule}) has the advantage of using only a single parameter, $\eta$.
	
	\subsection{Reducing Effect of Overlapping Speech} \label{ssec:overlapping}
	
	Since large-margin softmax losses only support a single target, overlapping speech will introduce additional complexities. Different methods can be used to combat this, such as just discarding the overlapping data, but instead a simpler way is introduced by utilising the flexibility of GLM-Softmax. 
	
	
	When the input is overlapping speech, the GLM-Softmax can be reduced to a modified softmax (equivalent to changing the loss parameters to $(m_1, m_2, m_3) = (1, 0, 0)$) to minimise the degrading effect overlapping speech has on the promotion of inter-class separability and intra-class compactness. 
	Therefore, within a single batch, two types of gradients will be computed: one based on modified softmax and the other, on GLM-Softmax. 
	
	
	\section{Experimental Setup} \label{sec:experimental}
	
	
	\subsection{Data Preparation}\label{ssec:data}
	
	All of the data preparation and model training was done using an extension of HTK version 3.5.1 and PyHTK \cite{HTK,PyHTK}.  The AMI meeting corpus \cite{AMI et al} was used for training with meetings recorded at four different sites. The full training set contains 135 meetings with 149 speakers recorded in total, of which, 10\% of the data for each speaker was used for the validation set.  Unlike \cite{P. Cyrta et al, G. Sun et al}, the full development (dev) and evaluation (eval) sets were used in scoring; the dev, but not the eval set, has a speaker overlap with the training set (see Table \ref{table:datastats}).
	\begin{table}[h!]
		\centering
		\vspace{-1mm}
		\begin{tabular}{c|c|c}
			\toprule
			Set & \# of Meetings & \# of Speakers \\
			\midrule
			Training & 135 & 149 \\
			Dev & 18 & 21 (2 in training) \\
			Eval & 16 & 16 (0 in training)\\
			\bottomrule
		\end{tabular}
		\vspace{-0.2cm}
		\caption{Number of speakers for each set.}
		\label{table:datastats}
	\end{table}
	\vspace{-2mm}
	
	The features used are 40-dimensional log-mel filter bank outputs obtained after applying \textit{BeamformIt} \cite{BeamForm et al} on the Multiple Distance Microphone data. 
	The filter bank outputs for each segment were normalised by the mean of the segment and the variance of the complete training set. The segments had the end silences trimmed, since the duration of the starting and ending silence varied significantly.

	\subsection{Model Specification and Baseline}\label{ssec:model}
	
	\begin{table*}[t]
		\centering
		\begin{tabular}{c|c|c|c}
			\toprule
			\multirow{2}{*}{Model} & Combined v1 & Combined v2 & Overlapping Speech Model \\
			& $(m_1, m_2, m_3) = (1.05, 0.08, 0.02)$ & $ (m_1, m_2, m_3) = (0.94, 0.20, 0.00)$ &$(m_1, m_2, m_3) = (1.045, 0.04, 0.05)$\\
			\midrule
			Dev & 13.4  & 13.5 & 13.2 \\ 
			Eval & 12.4  & 12.6 & 10.9 \\
			\bottomrule
		\end{tabular}
		\vspace{-0.2cm}
		\caption{SERs (\%) for models using all parameters in GLM-Softmax. The Overlapping Speech Model uses the approach of modifying the GLM-Softmax parameters for overlapping speech samples.}
		\label{table:multipleparameters}
	\end{table*}
	
\begin{figure}[h!]
		\centering
		\includegraphics[width=0.47\textwidth]{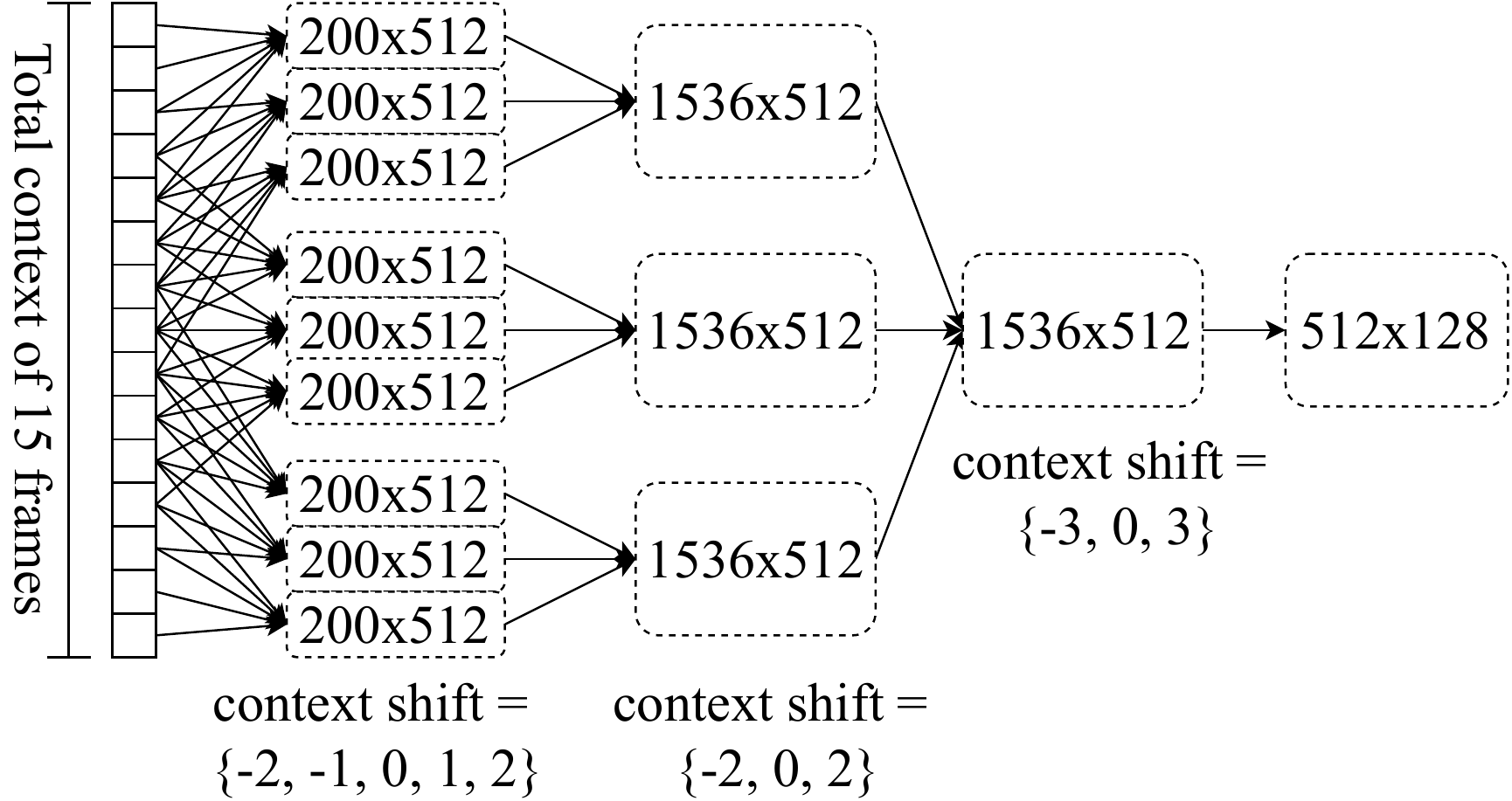}
		\vspace{-0.2cm}
		\caption{A visual representation of the frame level TDNN used. Each box represents a fully connected layer and multiple input arrows indicate concatenation. All activations functions used were ReLU.}
		\label{fig:frameleveltdnn}
	\end{figure}

	A frame level time delay neural network (TDNN), following the model in \cite{D. Snyder et al}, was used but with an output dimension of 128 (see Figure \ref{fig:frameleveltdnn}).
	
	The outputs of the frame level model on 200 consecutive frames were then combined in a self-attentive layer \cite{Y. Zhu et al}, with five attention heads. The output of the multi-head self-attentive layer, a $128 \times 5= 640$ dimensional vector, was then projected back to a 128-dimensional speaker embedding. During training the embedding was then input to a final fully connected layer which classified it into 149 speakers. The self-attentive layer was trained with an additional penalty term \cite{G. Sun et al} to diversify the annotation vectors for each head, in order to capture different aspects of a person’s speech. 
	
	The embeddings were extracted from a sliding context window of 2 seconds, with a 1 second shift. All activation functions were ReLU and the embeddings were extracted after the non-linearity in the last hidden layer. This was to retain as much information as possible about the angular nature of the embeddings, since the clustering process made use of cosine distances.
	
	The baseline used the model above, optimised for the modified softmax loss. Normalised weights, in the final layer, were achieved by renormalising to a magnitude of one after each update. The GLM-Softmax based models all used normalised weights with $\psi(\theta_t)$ from Eqn. (\ref{eq:gen_psi}), whilst models that tested the idea from Section \ref{ssec:overlapping}, had the GLM-Softmax reduced to a modified softmax loss for overlapping speech. Overlapping speech was trained on multiple times, each time with the label set to a different speaker in the input. 
	Furthermore, to stabilise training the update method proposed in Eqn. (\ref{eq:updateSchedule}), with $\eta = 1.25\times10^{-4}$ was used for all parameters. This value of $\eta$ was chosen together with a sliding context shift of one second, and a batch size of 200, since it was found that the loss parameters would be within 95\% of the desired value around 15 epochs and the models trained for approximately 20 epochs. 
	
	All models had the TDNN related parameters initialised by pre-training, while the remaining parameters, including the self-attentive layer, were randomly initialised. The pre-trained weights were generated from training the frame-level TDNN on the cross-entropy loss for 3 epochs with a context shift of 10 milliseconds to let the segment level models have a stable starting point. Afterwards, the self-attentive model was jointly trained with the frame-level TDNN.
	
	\subsection{Clustering and Diarisation}\label{ssec:diarisation}
	
	In order to make better use of the angular nature of the loss function, a modified spectral clustering approach from \cite{Q. Wang et al} based on cosine distance was employed. Instead of averaging embeddings from the same segment pre-clustering, the clustering process was first run with all embeddings, and based on the predicted speaker labels, speaker embedding centres were calculated. Then, the average speaker embedding (of each segment) was compared to the speaker centres, with the segment taking the label corresponding to the smallest cosine distance.
	
	The embeddings for clustering were produced from the same sliding context as in training, and the computed affinity matrix was based on cosine similarity. 
	The objective function and clustering parameters were tuned on the dev set for each model and then the results for both dev and eval are reported. Each meeting was assumed to have at least two speakers and therefore, the first eigenvalue of the processed affinity matrix was discarded when computing the number of clusters.
	
	Following previous work focused on the quality of speaker embeddings \cite{G. Sun et al, S.H. Yella et al, S.H. Yella et al v2}, AMI oracle segmentation was used and hence, only speaker error rates (SERs) are reported. For scoring, the NIST-RT evaluation setup was used with a 0.25 second collar.
	
	\section{Results} \label{sec:results}
	
	Table \ref{table:singleparameter} gives the SER results for the baseline and for each of the single parameter GLM-Softmax losses. It shows the loss outperforming the baseline with $m_1$ performing best on average. 
	
	\begin{table}[h!]
		\vspace{-1mm}
		\centering
		\begin{tabular}{c|c|ccc}
			\toprule
			\multirow{2}{*}{Model} & \multirow{2}{*}{Baseline} & \multicolumn{3}{c}{Single Parameter} \\
			&  & $m_1 = 1.10$ & $m_2 = 0.20$ & $m_3= 0.15$   \\
			\midrule
			Dev & 15.9 & 13.8 & 13.8 & 13.5  \\ 
			Eval & 18.3 & 12.8 & 13.9 & 13.8  \\
			\bottomrule
		\end{tabular}
		\vspace{-0.2cm}
		\caption{Baseline and single parameter GLM-Softmax SERs (\%).}
		\label{table:singleparameter}
	\end{table}
	\vspace{-1mm}
	
	For each of the parameters $m_1$, $m_2$, $m_3$ taken individually, there was a reduction in SER of 22.2\%, 19.0\% and 20.1\%, respectively. Here, $m_1$ was found to be best, which is in contrast to other works in speaker verification \cite{Y. Liu et al, Y. Yu et al} or in face verification \cite{H. Wang et al, J. Deng et al} that found $m_2$ or $m_3$ to be performing better. This may be explained by the approximation (see Eqn. (\ref{eq:approxpsi})) that these works have done which affects models using $m_1$ more. 
	
	
	The results for combining the parameters in unified way using GLM-Softmax are shown in Table \ref{table:multipleparameters}. The combined approach shows further improvements relative to the single parameter models. Combined v1 has a further 3.0\% SER relative reduction over $m_1 = 1.10$ (and 24.6\% relative over the baseline), with Combined v2 performing competitively. The notable property about v2 is that $m_1 < 1$ makes the training criteria harder as the model becomes more accurate, and $m_2 = 0.20$ ensures $\cos(\theta_t) \geqslant \psi(\theta_t)$. 
	
	For the Overlapping Speech Model (using the approach of changing parameters to $(m_1, m_2, m_3) = (1, 0, 0)$ for overlapping speech samples), since the AMI dataset includes a significant amount of overlapping speech, the best parameters to use are notably different from the conventional approach (of using just one set of parameters). Optimal parameters were found to be $(m_1, m_2, m_3) = (1.045, 0.04, 0.05)$ for this method, with further improvements of 6.6\% SER reduction relative to Combined v1, see Table \ref{table:multipleparameters}. Overall, the reduction in SER relative to the baseline is 17.0\% and 40.4\% for dev and eval respectively, giving an overall reduction of 29.5\%.

	\section{Conclusion}\label{sec:conclusion}
	
	This work has presented a general version of large-margin softmax without any approximations and a simple of method of stabilising training. A way to extend the use of the loss to overlapping speech has also been introduced. The proposed GLM-Softmax approach allows for generic use of all parameters in a unified manner without any approximation that could limit performance. 
	
	The use of the GLM-Softmax significantly improves the model performance, and combining the parameters in the loss further improves the performance to an overall 24.6\% SER reduction relative to the baseline. Moreover by changing the parameters of the loss when overlapping speech is present in training, further improvements can be noted achieving a 29.5\% SER reduction relative to baseline. This shows the power of GLM-Softmax, and that differentiating between normal and overlapping speech could lead to further improvements in diarisation if handled properly. 
	
	\vfill\pagebreak
	
	\renewcommand{\bibsection}{}
	
	\section{References}

\end{document}